\documentclass[3p]{elsarticle}

\usepackage{lineno}
\usepackage{hyperref}
\modulolinenumbers[1]

\journal{Physica A}

\usepackage{amsmath}
\usepackage{amsthm}
\usepackage{amssymb}
\usepackage{graphicx}
\usepackage{esint}
\usepackage{color}

\biboptions{numbers,sort&compress}

\begin{document}

\begin{frontmatter}

\title{On the quantum dynamics of Davydov solitons in protein $\alpha$-helices}

\author[address1]{Danko D. Georgiev\corref{mycorrespondingauthor}}
\ead{danko.georgiev@mail.bg}
\cortext[mycorrespondingauthor]{Corresponding author}

\author[address2]{James F. Glazebrook}
\ead{jfglazebrook@eiu.edu}

\address[address1]{Institute for Advanced Study, Varna, Bulgaria}
\address[address2]{Department of Mathematics and Computer Science, Eastern Illinois University, Charleston, IL, USA}

\begin{abstract}
The transport of energy inside protein $\alpha$-helices is studied
by deriving a system of quantum equations of motion from the Davydov
Hamiltonian with the use of the Schr\"{o}dinger equation and the generalized
Ehrenfest theorem. Numerically solving the system of quantum equations
of motion for different initial distributions of the amide I energy
over the peptide groups confirmed the generation of both moving or
stationary Davydov solitons. In this simulation the soliton generation, propagation,
and stability were found to be dependent on the symmetry of the exciton-phonon interaction Hamiltonian and the initial site of application of the exciton energy.
\end{abstract}

\begin{keyword}
Davydov soliton\sep numerical simulation\sep protein $\alpha$-helix\sep quantum dynamics\sep stability
\end{keyword}

\end{frontmatter}


\section{Introduction}

Proteins are the molecular engines of life that catalyze a myriad
of biochemical processes in living organisms \cite{Bu2011,Frauenfelder1991,GeorgievGlazebrook2018}.
The physical mechanisms for transport of energy inside proteins have
been of great interest to scientists ever since the discovery of the
protein secondary structure in 1951 by L.~Pauling and colleagues
\cite{Pauling1951}. In 1973, the pioneering work of A.~S.~Davydov
suggested that the energy released by adenosine triphosphate (ATP)
could be used to excite amide I oscillators in protein $\alpha$-helices,
which then interact with the induced distortion in a phonon lattice to
self-trap the amide I energy from dispersing thereby forming a waveform of `soliton' type \cite{Davydov1973,Davydov1973b,Davydov1976,Davydov1977,Davydov1979a,Davydov1979b,Davydov1982a}.
Davydov's model has been intensively studied by different research
teams \cite{Christiansen1990,Cruzeiro1997,Cruzeiro2009,Forner1991c,Forner1992a,Forner1996a,Forner1996b,Forner1997a,Forner1997b,Kerr1987,Kerr1990,Lawrence1986,Lawrence1987,Daniel2001,Daniel2002,Brizhik2004,Brizhik2006,Brizhik2010,Biswas2010,LeMesurier2012,Luo2017,Taghizadeh2017,Blyakhman2017,Brown1986,Brown1988,MacNeil1984,Scott1984,Cruzeiro1988},
but a number of biologically important questions remained unclear,
namely, whether the approximations in deriving the nonlinear Schr\"{o}dinger
equation \cite{Daniel2001,Biswas2010,Taghizadeh2017,Davydov1979a} for Davydov
solitons in proteins are justified for short protein $\alpha$-helices
that enter into the tertiary structure of most proteins, and whether
the system of coupled differential equations modeling the dynamics
of amide I excitons and the corresponding lattice distortions are
able to capture correctly the quantum nature of Davydov solitons, or
the approximations involved compel the system into a purely classical
regime \cite{Brown1986,Scott1984,MacNeil1984}.

Here, we investigate both of the above questions: Firstly, to an extent following \cite{Kerr1987,Kerr1990}, we derive
the quantum equations of motion resulting from solving the Schr\"{o}dinger
equation for Davydov's Hamiltonian, followed by an application of
the generalized Ehrenfest theorem for regulating the time evolution of quantum
expectation values. Then (in \S\ref{computational}) as a further mainstay of the present paper, we present a computational study of the dynamics of
the Davydov solitons in protein $\alpha$-helices, for the case of 40 peptide groups per $\alpha$-helix spine, including the size of the conformational
changes induced in the lattice, the velocity of soliton propagation
and its stability upon reflection from the ends of the protein $\alpha$-helix. The subsequent simulations, as depicted, reveal soliton dynamics for both symmetric and asymmetric Hamiltonians.

\section{Davydov's Hamiltonian}

\begin{figure}
\begin{centering}
\includegraphics[width=76mm]{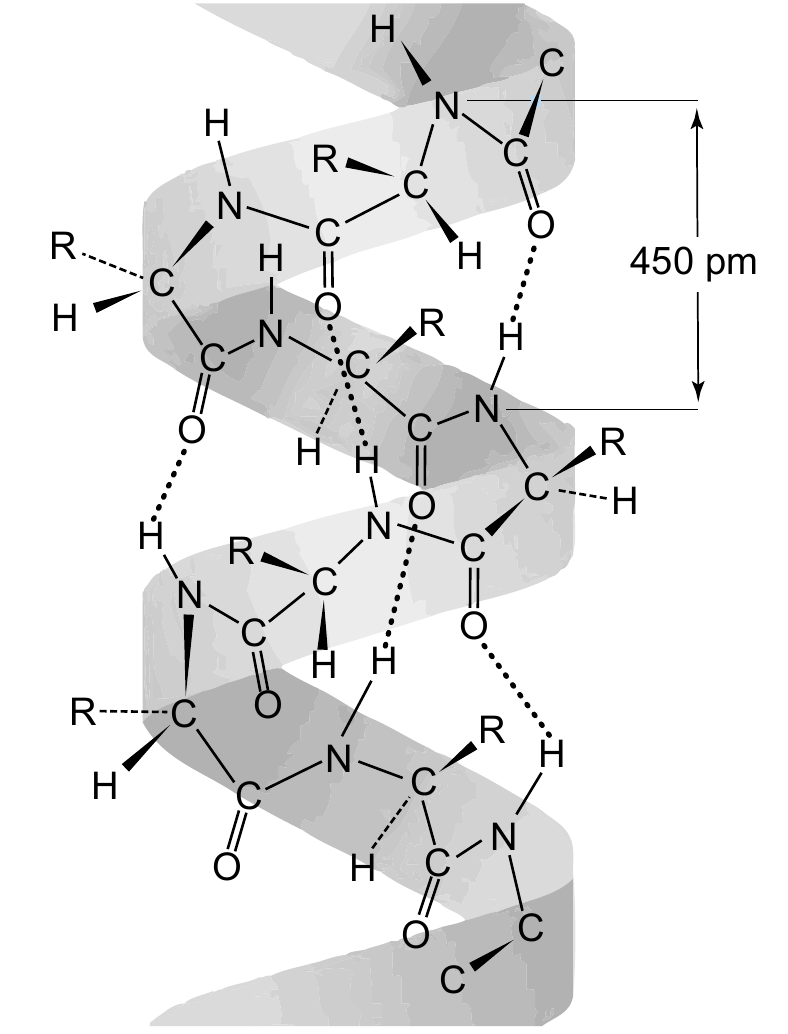}
\par\end{centering}

\caption{\label{fig:1}The protein $\alpha$-helix structure is stabilized
by 3 chains of hydrogen bonds referred to as $\alpha$-helix spines
that run parallel to the helical axis.}
\end{figure}

Geometrically, the protein $\alpha$-helix is a right-handed spiral
with 3.6 amino acid residues per turn, where the N--H group of an
amino acid forms a hydrogen bond with the C=O group of the amino acid
four residues positioned earlier in the polypeptide chain.
Three longitudinal chains of hydrogen bonds referred to as \mbox{$\alpha$-helix}
spines that run parallel to the helical axis stabilize the $\alpha$-helix
structure (Fig.~\ref{fig:1}). Each $\alpha$-helix spine consists of a chain of hydrogen
bonded peptide groups
\begin{equation*}
\cdots H-N-C\underset{J}{\underbrace{=O\cdots H-N-C}=}\underset{\chi}{\underbrace{O\cdots H}}-N-C=O\cdots
\end{equation*}
where $J$ is the interaction energy between two consecutive C=O groups,
and $\chi$ is the nonlinear coupling between the excited C=O group
and the distortion of the adjacent hydrogen bonds.

Focusing upon a single $\alpha$-helix spine, we decompose Davydov's
Hamiltonian as a sum of three separate Hamiltonians
\begin{equation}
\hat{H}=\hat{H}_{\textrm{ex}}+\hat{H}_{\textrm{ph}}+\hat{H}_{\textrm{int}}
\end{equation}
The above Hamiltonian is formally similar to the Fr\"{o}hlich--Holstein Hamiltonian which describes the interaction of electrons with a lattice \cite{Frohlich1952,Frohlich1954,Holstein1959a,Holstein1959b}.

The exciton (amide I; C=O stretching) energy operator is \cite{Davydov1982a,Scott1992}
\begin{equation}
\hat{H}_{\textrm{ex}}=\sum_{n}\left[E_{0}\hat{a}_{n}^{\dagger}\hat{a}_{n}-J\left(\hat{a}_{n}^{\dagger}\hat{a}_{n+1}+\hat{a}_{n}^{\dagger}\hat{a}_{n-1}\right)\right]
\end{equation}
where $\hat{a}_{n}^{\dagger}$ and $\hat{a}_{n}$ are respectively
the boson creation and annihilation operators for the amide~I oscillators,
$n\in\{1,2,\ldots,N\}$ counts the peptide groups along the spine,
$E_{0}=3.28\times10^{-20}$ J is the amide-I site energy, and $J=1.55\times10^{-22}$
J is the nearest neighbor dipole-dipole coupling energy along the
$\alpha$-helix spine \cite{Hyman1981,Nevskaya1976}.

The phonon energy operator is \cite{Davydov1982a,Kerr1987,Scott1992}
\begin{equation}
\hat{H}_{\textrm{ph}}=\frac{1}{2}\sum_{n}\left[\frac{\hat{p}_{n}^{2}}{M}+w\left(\hat{u}_{n+1}-\hat{u}_{n}\right)^{2}\right]
\end{equation}
where $M=1.9\times10^{-25}$ kg is the average mass of an amino acid,
$w$=13--19.5 N/m is an effective elasticity coefficient of the lattice
(the spring constant of the hydrogen bond)\cite{Scott1992}, $\hat{p}_{n}$ and $\hat{u}_{n}$
are respectively the momentum and position operators for longitudinal
displacement of an amino acid.

The exciton-phonon interaction operator is
\begin{equation}
\hat{H}_{\textrm{int}}=\sum_{n}\chi\left(\hat{u}_{n+1}+\left(\xi-1\right)\hat{u}_{n}-\xi\hat{u}_{n-1}\right)\hat{a}_{n}^{\dagger}\hat{a}_{n}
\end{equation}
where $\chi=30-62\times10^{-12}$ N is an anharmonic parameter arising
from the coupling between the quasiparticle (exciton) and the lattice
displacements (phonon) thereby parameterizing the strength of the
exciton-phonon interaction \cite{Scott1992}. {Here, we have introduced the parameter $\xi=[0,1]$
in order to characterize compactly the symmetry of the coupling between
the excited C=O group and the adjacent hydrogen bonds. For example,} Davydov's original
spatially symmetric model is obtained for $\xi=1$, where $\hat{a}_{n}^{\dagger}\hat{a}_{n}$
is coupled equally to $\hat{u}_{n+1}$ and $\hat{u}_{n-1}$ \cite{Davydov1982a,Kerr1987}.
A.~C.~Scott modified Davydov's model by opting for an asymmetric
coupling with $\xi=0$, motivated by the internal geometry of the
peptide units such that every unit has its C=O group immediately adjacent
to the next hydrogen bond in the chain; hence $\hat{a}_{n}^{\dagger}\hat{a}_{n}$
is coupled to $\hat{u}_{n+1}$, but not to $\hat{u}_{n-1}$ \cite{Scott1992}.
{Our generalized interaction Hamiltonian can be brought into the form \cite{Luo2017} given by Luo and Piette 
\begin{equation}
\hat{H}_{\textrm{int}}=\sum_{n}\bar{\chi}\left[(1+\beta)\hat{u}_{n+1}-2\beta\hat{u}_{n}-(1-\beta)\hat{u}_{n-1}\right]\hat{a}_{n}^{\dagger}\hat{a}_{n}
\end{equation}
where $\beta=\frac{1-\xi}{1+\xi}$ and $\bar{\chi}=\chi\frac{1+\xi}{2}$, or in terms of left $\chi_l$ and right $\chi_r$ coupling parameters, $\beta=\frac{\chi_r-\chi_l}{\chi_r+\chi_l}$, $\bar{\chi}=\frac{\chi_r+\chi_l}{2}$, $\chi=\chi_r$ and $\xi=\frac{\chi_l}{\chi_r}$.}

\section{Equations of motion for Davydov's Hamiltonian}

Here we show how the quantum equations of motion for Davydov's Hamiltonian can be
derived using only the Schr\"{o}dinger equation together with the generalized Ehrenfest
theorem \cite{Kerr1987,Kerr1990}.

First, we introduce an ansatz state vector whose time evolution is
assumed to be approximately the same as that of the exact
state vector. In the literature, Davydov introduced two different
possible state vectors called $|D_{1}\rangle$ or $|D_{2}\rangle$ ans\"{a}tze.
{ Because Davydov's Hamiltonian describing the transport of acoustic polarons is mathematically similar to the Holstein Hamiltonian describing optical polarons, the two Davydov ans\"{a}tze and the mathematical techniques employed in the analysis of Davydov's soliton are also often used to study polaron dynamics in molecular rings and other aggregates \cite{Sun2010,Sun2014}.}
Here, we work with the second of Davydov's ansatz state vectors, which
has the form \cite{Davydov1982a}:
\begin{equation}
|D_{2}(t)\rangle=|a\rangle|b\rangle;\quad
|a\rangle=\sum_{n}a_{n}(t)\hat{a}_{n}^{\dagger}|0_{\textrm{ex}}\rangle;\quad
|b\rangle=e^{-\frac{\imath}{\hbar}\sum_{j}\left(b_{j}(t)\hat{p}_{j}-c_{j}(t)\hat{u}_{j}\right)}
|0_{\textrm{ph}}\rangle\label{eq:D2}
\end{equation}
Normalization of the $|D_{2}\rangle$ ansatz implies
$\langle D_{2}|D_{2}\rangle=\sum_{n}\left|a_{n}\right|^{2}=1$,
where $\left|a_{n}\right|^{2}$ is the probability for finding the
amide I quantum exciton at the $n$th site.

With the use of the Hadamard lemma \cite[p.~143]{Bowen2015}, the expectation values for $\hat{u}_{n}$ and $\hat{p}_{n}$ are found to be
\begin{equation}
b_{n}(t) = \langle D_{2}(t)|\hat{u}_{n}|D_{2}(t)\rangle;\qquad
c_{n}(t) = \langle D_{2}(t)|\hat{p}_{n}|D_{2}(t)\rangle\label{eq:<c>}
\end{equation}
{If the $|D_{2}(t)\rangle$ ansatz approximates well the exact solution of the Schr\"{o}dinger equation (see \ref{appB}), then its temporal evolution will be}
\begin{equation}
\imath\hbar\frac{d}{dt}|D_{2}(t)\rangle=\hat{H}|D_{2}(t)\rangle\label{eq:Schrodinger}
\end{equation}
and we can use the generalized Ehrenfest theorem (see \ref{appA})
for the time dynamics of the expectation values \eqref{eq:<c>}, namely
\begin{equation}
\frac{d}{dt}b_{n} = \frac{1}{\imath\hbar}\langle\left[\hat{u}_{n},\hat{H}\right]\rangle; \qquad
\frac{d}{dt}c_{n} = \frac{1}{\imath\hbar}\langle\left[\hat{p}_{n},\hat{H}\right]\rangle\label{eq:Eh-c}
\end{equation}
For the above commutators, we obtain
\begin{eqnarray}
\left[\hat{u}_{n},\hat{H}\right]
& = & \imath\hbar\frac{\hat{p}_{n}}{M}\label{eq:com-u-H}\\
\left[\hat{p}_{n},\hat{H}\right]
& = & \imath\hbar w\Big(\hat{u}_{n+1}-2\hat{u}_{n}+\hat{u}_{n-1})-\imath\hbar\chi\Big(\hat{a}_{n-1}^{\dagger}\hat{a}_{n-1}+(\xi-1)\hat{a}_{n}^{\dagger}\hat{a}_{n}-\xi\hat{a}_{n+1}^{\dagger}\hat{a}_{n+1}\Big)\label{eq:com-p-H}
\end{eqnarray}
From \eqref{eq:Eh-c} and \eqref{eq:com-u-H}, we also have
\begin{equation}
\frac{d}{dt}b_{n} =\frac{c_{n}}{M};\qquad
\frac{d}{dt}c_{n} = M\frac{d^{2}}{dt^{2}}b_{n}
\label{eq:22}
\end{equation}
From \eqref{eq:Eh-c} and \eqref{eq:com-p-H}, together with
\eqref{eq:<c>} and \ref{eq:22}, we obtain one of Davydov's equations for the phonon displacements $b_{n}(t)$
from the corresponding equilibrium positions
\begin{equation}
M\frac{d^{2}}{dt^{2}}b_{n}=w\Big(b_{n+1}-2b_{n}+b_{n-1})+\chi\Big(\xi\left|a_{n+1}\right|^{2}+(1-\xi)\left|a_{n}\right|^{2}-\left|a_{n-1}\right|^{2}\Big)\label{eq:Davydov-1b}
\end{equation}

The equation for the amide I probability amplitudes $a_{n}(t)$ can
be derived by differentiating the $|D_{2}(t)\rangle$ ansatz.
After application of the Baker--Campbell--Hausdorff formula \cite{VanBrunt2015},
the total time derivative of the Davydov ansatz $|D_{2}(t)\rangle$
is found to be \cite{Kerr1987}
\begin{equation}
\imath\hbar\frac{d}{dt}|D_{2}(t)\rangle=\imath\hbar\sum_{n}\frac{da_{n}}{dt}\hat{a}_{n}^{\dagger}|0_{\textrm{ex}}\rangle|b\rangle+|a\rangle\sum_{j}\left(\frac{db_{j}}{dt}\hat{p}_{j}-\frac{dc_{j}}{dt}\hat{u}_{j}+\frac{1}{2}\left(b_{j}\frac{dc_{j}}{dt}-\frac{db_{j}}{dt}c_{j}\right)\right)|b\rangle\label{eq:dD2/dt}
\end{equation}
Next, we calculate the terms on right-hand side of the Schr\"{o}dinger equation as follows
\begin{eqnarray}
\hat{H}_{\textrm{ex}}|D_{2}(t)\rangle
& = & \sum_{n}\Bigg[E_{0}a_{n}-J\left(a_{n+1}+a_{n-1}\right)\Bigg]\hat{a}_{n}^{\dagger}|0_{\textrm{ex}}\rangle|b\rangle\label{eq:HexD2}\\
\hat{H}_{\textrm{ph}}|D_{2}(t)\rangle & = & \sum_{n}\hat{H}_{\textrm{ph}}\hat{a}_{n}^{\dagger}|0_{\textrm{ex}}\rangle|b\rangle\label{eq:HphD2}\\
\hat{H}_{\textrm{int}}|D_{2}(t)\rangle
& = & \chi\sum_{n}a_{n}(t)\left(\hat{u}_{n+1}+(\xi-1)\hat{u}_{n}-\xi\hat{u}_{n-1}\right)\hat{a}_{n}^{\dagger}|0_{\textrm{ex}}\rangle|b\rangle\label{eq:HintD2}
\end{eqnarray}
Then we use the Schr\"{o}dinger equation to combine \eqref{eq:dD2/dt}, \eqref{eq:HexD2}, \eqref{eq:HphD2} and \eqref{eq:HintD2},
and after taking the inner product with $\langle b|\langle0_{\textrm{ex}}|\hat{a}_{n}$, we obtain
\begin{equation}
\imath\hbar\frac{da_{n}}{dt} = \Bigg[E_{0}+W(t)-\frac{1}{2}\sum_{j}\left(\frac{db_{j}}{dt}c_{j}-b_{j}\frac{dc_{j}}{dt}\right)+\chi\left(b_{n+1}+(\xi-1)b_{n}-\xi b_{n-1}\right)\Bigg]a_{n}
 -J\left(a_{n+1}+a_{n-1}\right)\label{eq:Davydov-2a}
\end{equation}
where the expectation value of the phonon energy has been written as
$W(t)=\langle D_{2}|\hat{H}_{\textrm{ph}}|D_{2}\rangle$.

The first three terms in \ref{eq:Davydov-2a} are global for all $a_{n}$,
namely
\begin{equation}
\forall a_{n}:\qquad\gamma(t)=E_{0}+W(t)-\frac{1}{2}\sum_{j}\left(\frac{db_{j}}{dt}c_{j}-b_{j}\frac{dc_{j}}{dt}\right)
\end{equation}
Furthermore, because all terms in $\gamma(t)$ are real, a global phase change
on the quantum probability amplitudes, namely $a_{n}\to\bar{a}_{n}e^{-\frac{\imath}{\hbar}\int\gamma(t)dt}$,
will not change the quantum probabilities for the amide I oscillators
\begin{equation}
|a_{n}|^{2}=e^{+\frac{\imath}{\hbar}\int\gamma(t)dt}\bar{a}_{n}^{*}\bar{a}_{n}e^{-\frac{\imath}{\hbar}\int\gamma(t)dt}=|\bar{a}_{n}|^{2}
\end{equation}
After the transformation and re-labeling $\bar{a}_{n}$ to $a_n$, the system of Davydov equations becomes
\begin{eqnarray}
\imath\hbar\frac{da_{n}}{dt} & = & \chi\left[b_{n+1}+(\xi-1)b_{n}-\xi b_{n-1}\right]a_{n}-J\left(a_{n+1}+a_{n-1}\right)\\
M\frac{d^{2}}{dt^{2}}b_{n} & = & w\Big(b_{n-1}-2b_{n}+b_{n+1})-\chi\Big(\left|a_{n-1}\right|^{2}+(\xi-1)\left|a_{n}\right|^{2}-\xi\left|a_{n+1}\right|^{2}\Big)
\end{eqnarray}

\section{\label{computational}Computational study}

\subsection{Model parameters}

We have numerically integrated the system of Davydov equations
for an $18$-nm-long protein $\alpha$-helix with $n=40$ peptide
groups per $\alpha$-helix spine with initially unperturbed lattice of hydrogen bonds
and different initial conditions for spreading of the amide~I excitation.
Because counting of amino acid residues in the protein primary structure starts, by
definition, at the N-end of the protein, the index $n$ denoting the
peptide groups along the $\alpha$-helix spine is minimal at the N-end
and maximal at the C-end of the $\alpha$-helix.
For improving comparability with previous works \cite{Cruzeiro1988,MacNeil1984,Scott1984}, in the present simulations we have set $w=13$ N/m and have varied the values of $\bar{\chi}$ and $\xi$. The error of deviation from the Schr\"{o}dinger equation due to the use of $|D_2(t)\rangle$ ansatz is analyzed in \ref{appB} and is numerically found to be negligible in comparison to
the absolute value of the gauge transformed soliton energy $|E|= |E_{\textrm{sol}} - E_0 - W_0|$.

\subsection{Effect of initial exciton spreading}

\begin{figure}
\begin{centering}
\includegraphics[width=160mm]{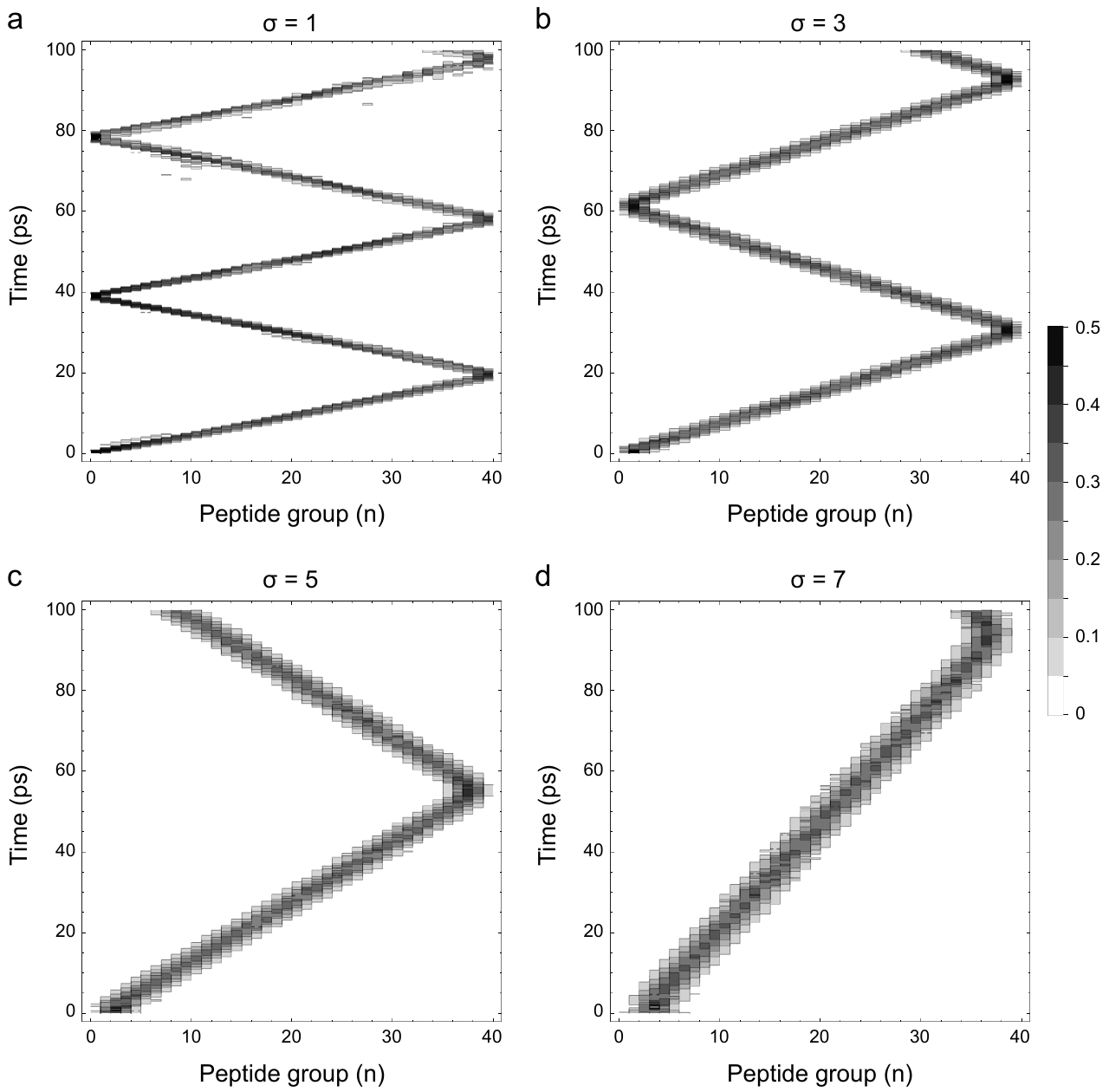}
\par\end{centering}
\caption{\label{fig:2}Soliton dynamics visualized through $|a_n|^2$ for the symmetric Hamiltonian $\xi=1$
at supra-threshold value $\bar{\chi}=35$~pN for different initial Gaussian step distributions $\sigma$ of amide~I energy
over 1, 3, 5 or 7 peptide groups starting from the N-end
of an $\alpha$-helix spine composed of 40 peptide groups during a period of 100 ps.}
\end{figure}

In the continuum approximation, the Davydov soliton is a $\textrm{sech}^2$-shaped distribution of amide~I energy,
which is self-trapped by the induced distortion in the lattice.
To investigate how the initial spreading of amide~I energy affects the dynamics of Davydov solitons in the discrete case, we have compared different initial Gaussian step distributions $\sigma$ of amide I energy spread over 1, 3,
5 or 7 peptide groups such that the corresponding distributions of
$|a_{n}|^{2}$ were computed to be: $\{1\}$, $\{0.244,0.512,0.244\},$ $\{0.099,0.24,0.322,0.24,0.099\}$,
$\{0.059,0.126,0.199,0.232,0.199,0.126,0.059\}$).
Simulations with $\xi=1$ and $\bar{\chi}=35$~pN showed that the larger initial spreading of amide~I energy generates solitons with lower energies and lower speeds:
for $\sigma=1$ the soliton energy was $E=0$ and the soliton velocity~$v=947$~m/s (Fig.~\ref{fig:2}a);
for $\sigma=3$ : $E =-1.41 J$ and $v=590$~m/s (Fig.~\ref{fig:2}b);
for $\sigma=5$ : $E =-1.73 J$ and $v=324$~m/s (Fig.~\ref{fig:2}c); and
for $\sigma=7$ : $E =-1.84 J$ and $v=189$~m/s (Fig.~\ref{fig:2}d).
Thus, the initial spreading of amide~I energy appears to stabilize the Davydov solitons by lowering of the soliton energy.

For simulations with initially unperturbed lattice, $b_n(0)=0$ and $\frac{db_{n}(0)}{dt}=0$, the soliton energy does not depend on $\bar{\chi}$ and $\xi$, but only on the initial spread of the amide~I energy (for details see Eq.~\ref{eq:solE} and the following discussion in \ref{appB}).
All subsequent simulations were performed with initial Gaussian step distribution $\sigma=3$, which was computationally confirmed to generate Davydov solitons with the same energy $E =-1.41 J$.

\subsection{Effect of increased exciton-phonon coupling}

\begin{figure}
\begin{centering}
\includegraphics[width=160mm]{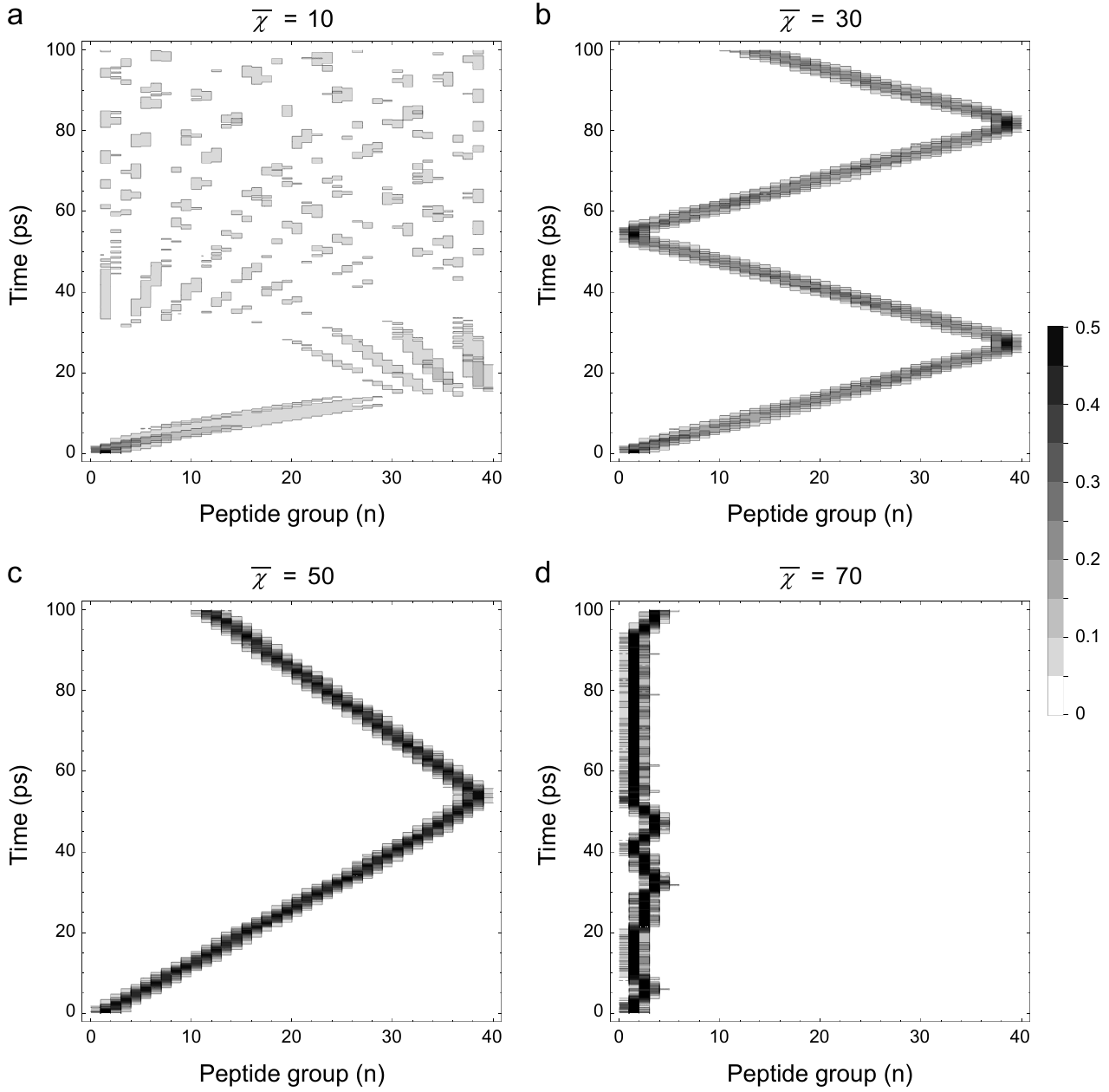}
\par\end{centering}
\caption{\label{fig:3}Soliton generation and pinning visualized through $|a_n|^2$ at different values of $\bar{\chi}$ for the symmetric Hamiltonian $\xi=1$
with an initial Gaussian step distribution $\sigma$ of amide~I energy over 3 peptide groups starting from the N-end
of an $\alpha$-helix spine composed of 40~peptide groups during a period of 100 ps.}
\end{figure}

To investigate how the change of the exciton-phonon coupling $\bar{\chi}$ affects the dynamics of Davydov solitons, we have varied $\bar{\chi}$ for fixed $\xi=1$ and initial Gaussian step distribution $\sigma=3$ of amide~I energy applied at the N-end of the $\alpha$-helix spine.
For $\bar{\chi}=10$~pN, the exciton energy interacted only weakly with the lattice and was rapidly dispersed along the $\alpha$-helix spine (Fig.~\ref{fig:3}a, Video~1).
For $\bar{\chi}=30$~pN, the exciton energy was self-trapped by an induced distortion in the lattice and the soliton did not disperse even after reflection from the end of the $\alpha$-helix spine (Fig.~\ref{fig:3}b, Video~2). The velocity of the moving soliton was $v=679$~m/s.
For $\bar{\chi}=50$~pN, the self-trapping was more pronounced and the soliton velocity was lower, $v=340$~m/s (Fig.~\ref{fig:3}c, Video~3).
For $\bar{\chi}=70$~pN, the self-trapping was so strong that it pinned the soliton and prevented it from moving along the $\alpha$-helix spine (Fig.~\ref{fig:3}d, Video~4).
Thus, the exciton-phonon coupling parameter $\bar{\chi}$ exhibits two distinct thresholds, a lower one for the formation of moving solitons and a higher one for soliton pinning.

\subsection{Effect of exciton-phonon interaction symmetry}

\begin{figure}
\begin{centering}
\includegraphics[width=80mm]{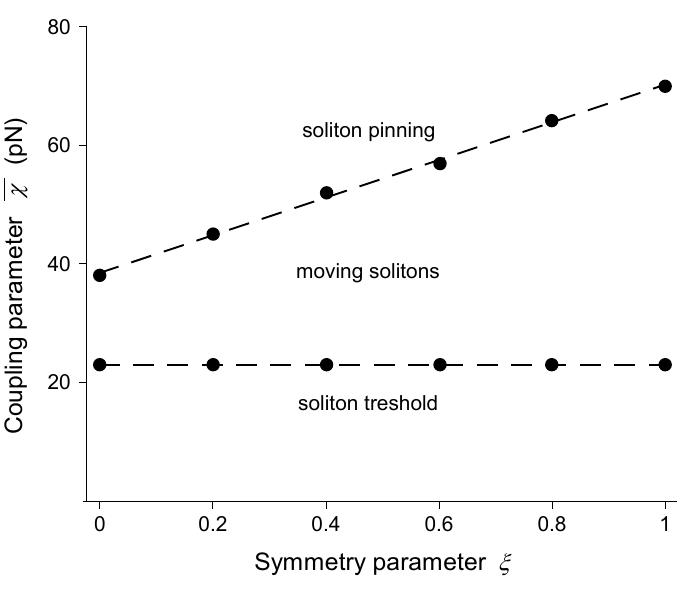}
\par\end{centering}
\caption{\label{fig:4}Thresholds for soliton generation and pinning at different values of $\bar{\chi}$ for different values of the symmetry parameter~$\xi$ with an initial Gaussian step distribution $\sigma$ of amide~I energy over 3 peptide groups starting from the N-end
of an $\alpha$-helix spine composed of 40~peptide groups during a period of 100 ps. Linear trend lines (dashed lines) are fitted to the values observed in the simulations (black circles).}
\end{figure}

To investigate how the exciton-phonon interaction symmetry $\xi$ affects the dynamics of Davydov solitons, we have varied both $\xi$ and $\bar{\chi}$ for fixed initial distribution $\sigma=3$. If the amide~I energy was applied at the N-end of the $\alpha$-helix spine, the threshold for generation of moving solitons was $\bar{\chi}=23$~pN for any $\xi$ (Fig.~\ref{fig:4}). This threshold is consistent with the one reported in previous simulations for $\xi=0$ \cite{Kenkre1994,Cruzeiro1988}.
Gaussians with larger initial spread, exhibited slightly lower thresholds for soliton generation: $\bar{\chi}=22$~pN for $\sigma=5$ and $\bar{\chi}=21$~pN for $\sigma=7$. Thus, launching of the Davydov solitons is modestly assisted by the initial spreading of the amide~I energy.

While the threshold for soliton generation was the same for all $\xi$,
the threshold for soliton pinning was strongly dependent on $\xi$.
With fixed $\sigma=3$, the thresholds for soliton pinning were:
for $\xi=0$ : $\bar{\chi}=38$~pN,
for $\xi=0.2$ : $\bar{\chi}=45$~pN,
for $\xi=0.4$ : $\bar{\chi}=52$~pN,
for $\xi=0.6$ : $\bar{\chi}=57$~pN,
for $\xi=0.8$ : $\bar{\chi}=64$~pN, and
for $\xi=1$ : $\bar{\chi}=70$~pN (Fig.~\ref{fig:4}).
Thus, the symmetry of the exciton-phonon interaction Hamiltonian increases the dynamic range of $\bar{\chi}$ values for which the generated solitons are able to move along the protein $\alpha$-helix.

\begin{figure}
\begin{centering}
\includegraphics[width=160mm]{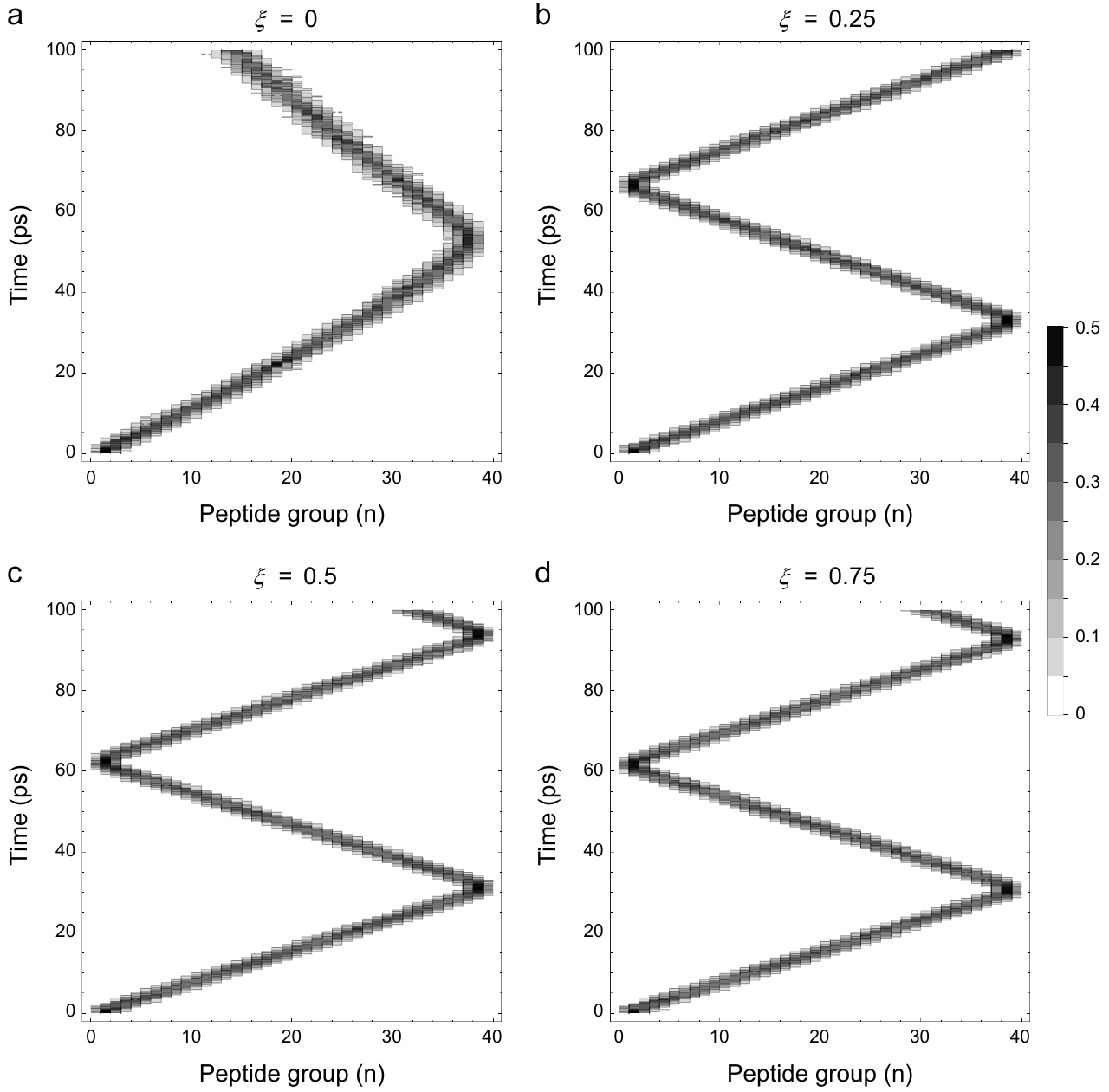}
\par\end{centering}
\caption{\label{fig:5}Soliton dynamics visualized through $|a_n|^2$ for different values of the symmetry parameter~$\xi$
at supra-threshold value $\bar{\chi}=35$~pN with an initial Gaussian step distribution $\sigma$ of amide~I energy over 3 peptide groups starting from the N-end of an $\alpha$-helix spine composed of 40 peptide groups during a period of 100 ps.}
\end{figure}

Higher values for the symmetry parameter $\xi$ also led to faster moving solitons.
With fixed $\sigma=3$ and $\bar{\chi}=35$~pN, the soliton velocities were:
for $\xi=0$ : $v=340$~m/s (Fig.~\ref{fig:5}a),
for $\xi=0.25$ : $v=545$~m/s (Fig.~\ref{fig:5}b),
for $\xi=0.5$ : $v=571$~m/s (Fig.~\ref{fig:5}c),
for $\xi=0.75$ : $v=586$~m/s (Fig.~\ref{fig:5}d),
for $\xi=1$ : $v=590$~m/s (Fig.~\ref{fig:2}b).
Interestingly, the presence of even a small amount of symmetry ($\xi=0.25$) made the soliton properties closer to the fully symmetric case ($\xi=1$) rather than the fully asymmetric case ($\xi=0$).

\begin{figure}
\begin{centering}
\includegraphics[width=160mm]{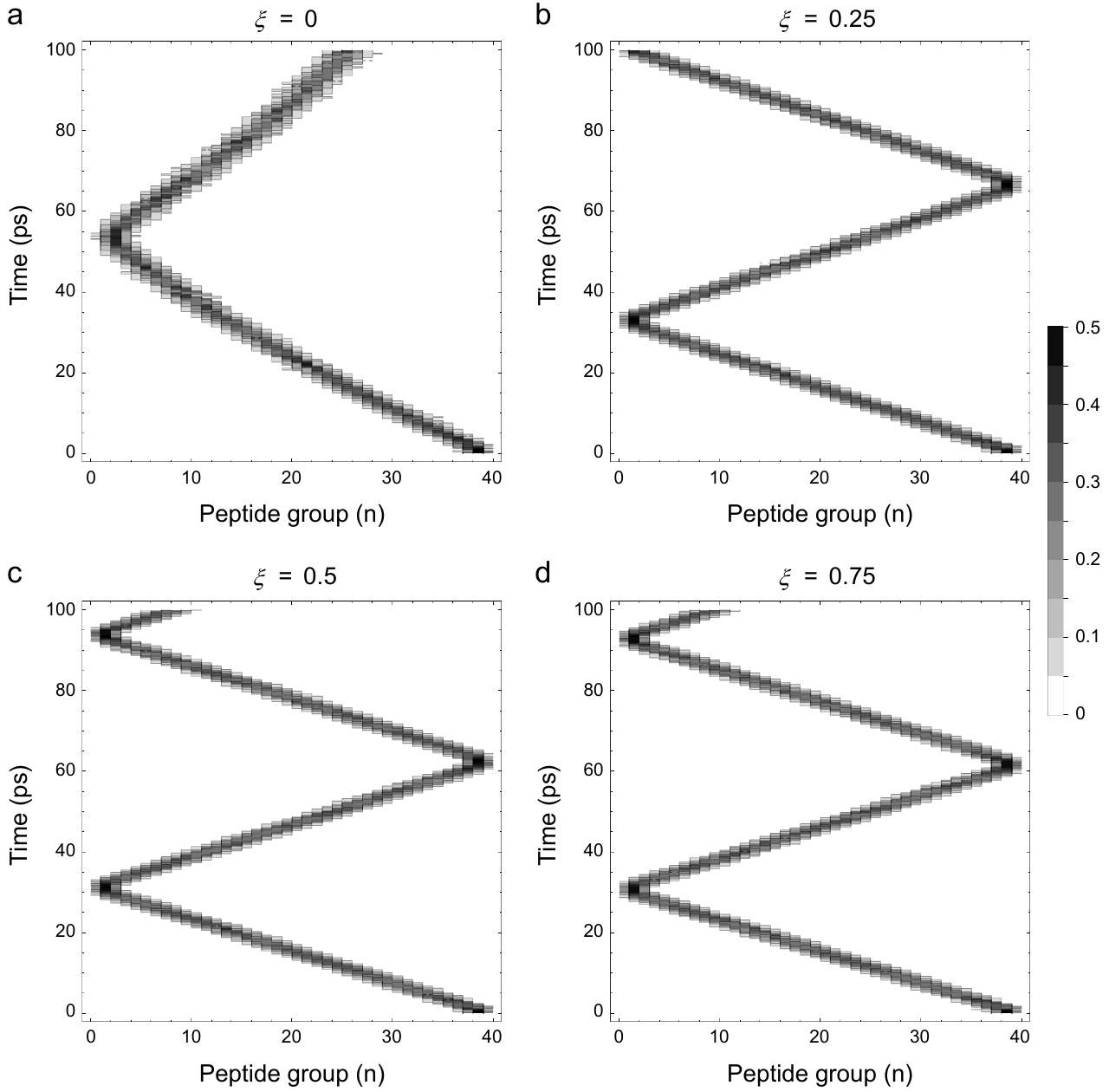}
\par\end{centering}
\caption{\label{fig:6}Soliton dynamics visualized through $|a_n|^2$ for different values of the symmetry parameter~$\xi$
at supra-threshold value $\bar{\chi}=35$~pN with an initial Gaussian step distribution $\sigma$ of amide~I energy over 3 peptide groups starting from the C-end of an $\alpha$-helix spine composed of 40 peptide groups during a period of 100 ps.}
\end{figure}

We have further compared the launching of Davydov solitons from the N-end versus the C-end of the protein $\alpha$-helix. The results for each $\xi$ value were almost identical mirror reflections (Figs.~\ref{fig:5} and \ref{fig:6}). The mirror symmetry was exact for $\xi=1$, and very slightly violated for $\xi<1$.

\begin{figure}
\begin{centering}
\includegraphics[width=160mm]{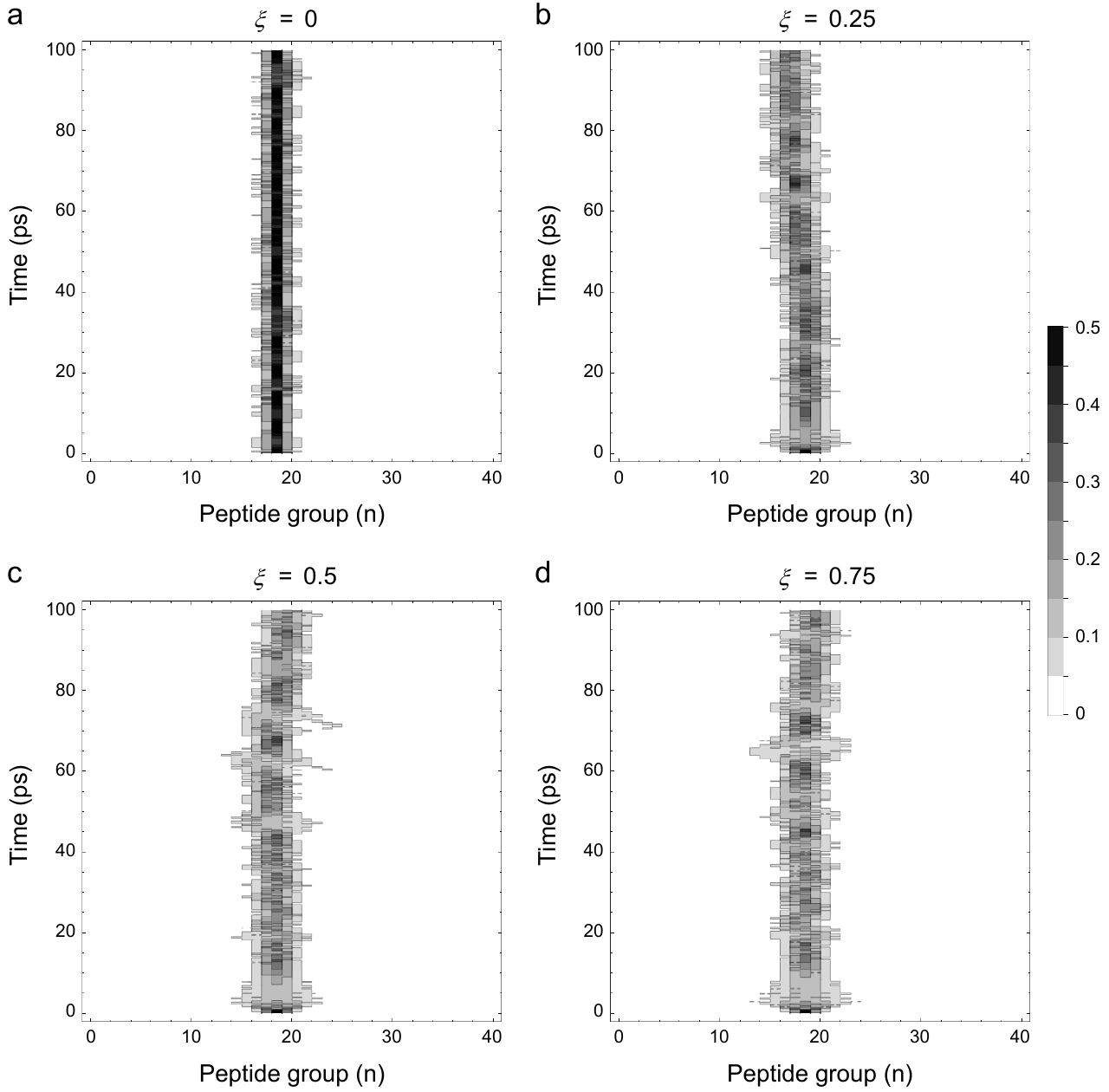}
\par\end{centering}
\caption{\label{fig:7}Soliton dynamics visualized through $|a_n|^2$ for different values of the symmetry parameter~$\xi$
at supra-threshold value $\bar{\chi}=40$~pN with an initial Gaussian step distribution $\sigma$ of amide~I energy over 3 peptide groups starting in the middle of an $\alpha$-helix spine composed of 40 peptide groups during a period of 100 ps.}
\end{figure}

Applying the amide~I energy in the middle of the protein $\alpha$-helix resulted in the generation of pinned solitons (Fig.~\ref{fig:7}).
The threshold for soliton generation was substantially higher in comparison with launching of solitons from the $\alpha$-helix ends and exhibited slight $\xi$-dependence.
The thresholds for soliton generation were:
for $\xi=0$ : $\bar{\chi}=34$~pN,
for $\xi=0.25$ : $\bar{\chi}=36$~pN,
for $\xi=0.5$ : $\bar{\chi}=37$~pN, and
for $\xi\geq0.75$ : $\bar{\chi}=37$~pN.
Thus, the exciton-phonon interaction asymmetry favors soliton pinning at lower values of $\bar{\chi}$ for all possible sites of application of the amide~I energy in the $\alpha$-helix.

\subsection{Effect of periodic boundary conditions}

\begin{figure}
\begin{centering}
\includegraphics[width=160mm]{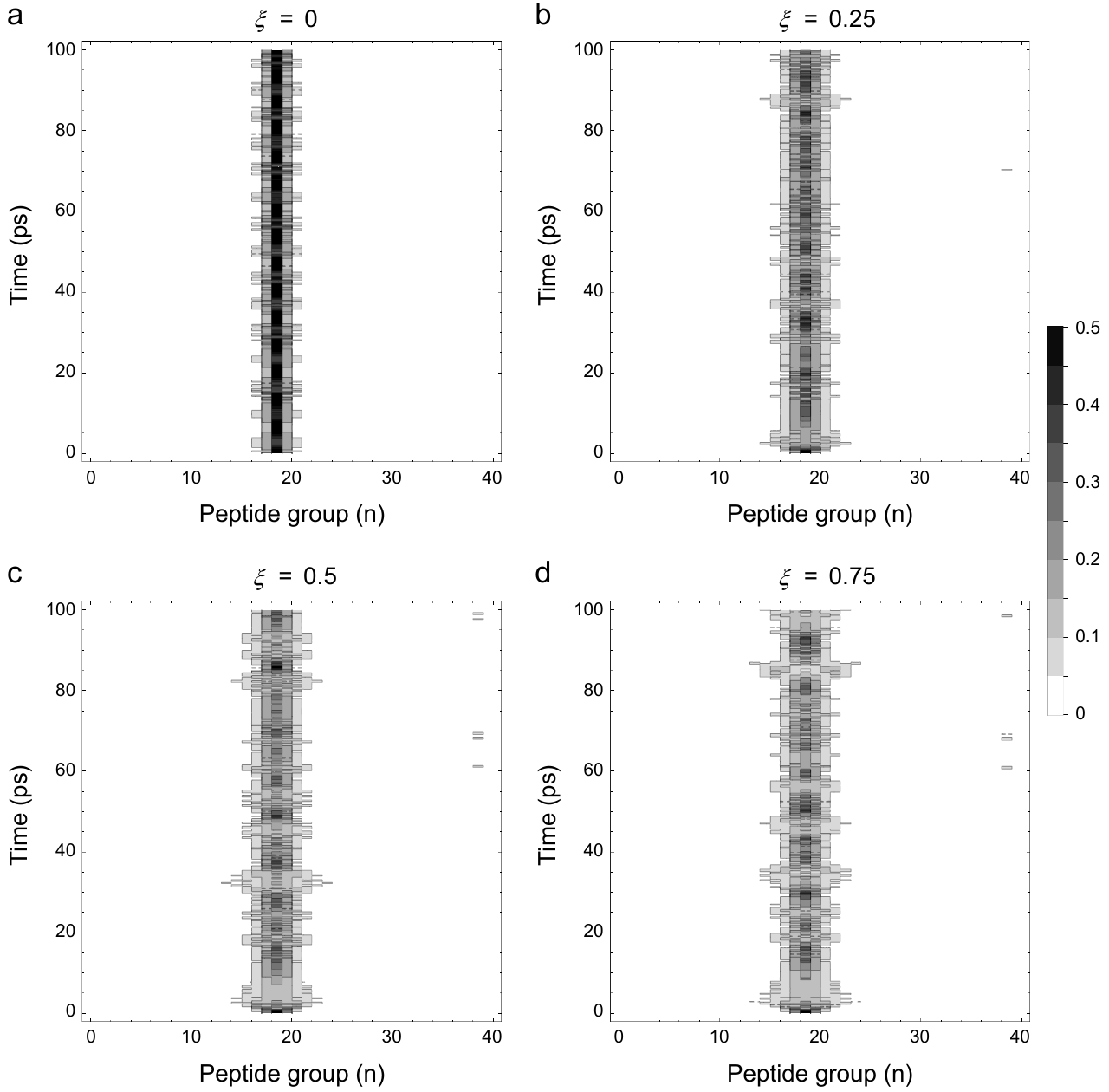}
\par\end{centering}
\caption{\label{fig:8}Soliton dynamics visualized through $|a_n|^2$ with imposed periodic boundary conditions (effectively circular $\alpha$-helix spine) for different values of the symmetry parameter~$\xi$
at supra-threshold value $\bar{\chi}=40$~pN with an initial Gaussian step distribution $\sigma$ of amide~I energy over 3 peptide groups in the $\alpha$-helix spine composed of 40 peptide groups during a period of 100 ps.}
\end{figure}

Periodic boundary conditions are often used in quantum chemistry calculations \cite{Kolev2013,Kolev2018} even though they effectively create circular lattices. In a circular $\alpha$-helix, the resulting solitons are pinned (Fig.~\ref{fig:8}) similarly to the case when the amide~I energy is applied in the middle of the $\alpha$-helix (Fig.~\ref{fig:7}).
For the periodic case, the thresholds for soliton generation were slightly higher:
for $\xi=0$ : $\bar{\chi}=34$~pN,
for $\xi=0.25$ : $\bar{\chi}=37$~pN,
for $\xi=0.5$ : $\bar{\chi}=38$~pN, and
for $\xi\geq0.75$ : $\bar{\chi}=38$~pN.
Thus, using periodic boundary conditions could be useful in determining the threshold for generation of solitons in the middle of the $\alpha$-helix or analyzing the soliton stability, but is inadequate for studying moving solitons along the $\alpha$-helix.

\begin{figure}
\begin{centering}
\includegraphics[width=80mm]{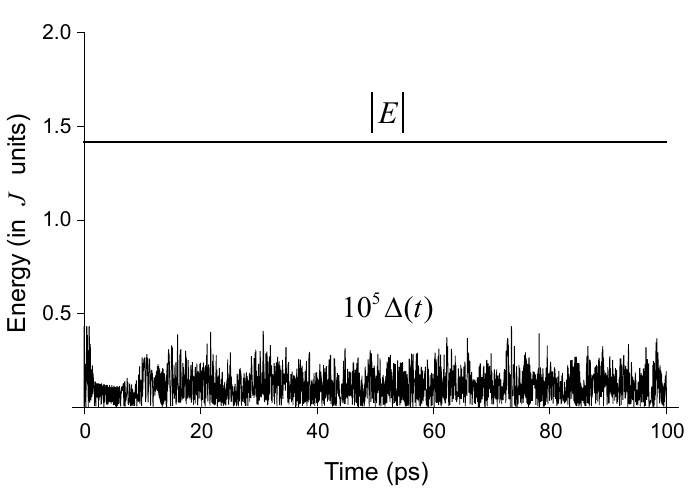}
\par\end{centering}
\caption{\label{fig:9}Numerical quantification of the error in the simulated quantum dynamics due to the use of $|D_2(t)\rangle$ ansatz for the case shown in Fig.~\ref{fig:2}b with $\xi=1$, $\bar{\chi}=35$~pN and $\sigma=3$. The error $\Delta(t)$ for deviation from the Schr\"{o}dinger equation is over $10^5$ times smaller than the absolute value of the gauge transformed soliton energy $|E|= |E_{\textrm{sol}} - E_0 - W_0|$.}
\end{figure}

\section{Discussion}

The transport of energy in proteins is an important problem, which
requires a detailed analysis of a number of physical parameters. The
discreteness of the peptide groups within the relatively short lengths
of average $\alpha$-helices entering in the protein tertiary structure
precludes the long-wavelength/continuum approximation \cite{Davydov1973b,Davydov1976}
with subsequent derivation of the nonlinear Schr\"{o}dinger equation, which
is known to have soliton solutions. With the use of the Schr\"{o}dinger
equation and the generalized Ehrenfest theorem, however, we have shown
that it is possible to derive a system of quantum equations of motion
that could be solved through numerical integration for arbitrary initial
conditions.
We have also addressed previous concerns \cite{Brown1986,Brown1988} in regard to the usage of $|D_2(t)\rangle$ ansatz for determining the full quantum dynamics by showing that the error of deviation from the Schr\"{o}dinger equation is non-zero, but negligible in comparison to the soliton energy.

Different initial Gaussian step distributions $\sigma$ of amide I energy exhibited self-trapping due to an associated kink in the lattice
of hydrogen bonds and moved as a solitary quasiparticle (viz. Davydov soliton). The symmetry of the Hamiltonian (set by the parameter $\xi$)
{increased the dynamic range of $\bar{\chi}$ values for which the solitons were moving and also increased the soliton velocities for each individual value of $\bar{\chi}$. Furthermore, the presence of even a small amount of symmetry, makes the soliton properties closer to the fully symmetric case rather than the fully asymmetric case. Thus, any non-zero value of the symmetry parameter $\xi$ is beneficial for the possible existence of moving Davydov solitons in protein $\alpha$-helices.}

Our results, as visually presented (Figs.~1--9; Videos~1--4) in the computational study (\S\ref{computational}), show that Davydov solitons can be generated even when the long-wavelength/continuum approximation cannot be used for the derivation of the nonlinear Schr\"{o}dinger equation in possession of soliton solutions. In such cases, the symmetry of the Hamiltonian is of great importance for the mobility of the generated solitons and the velocity of their propagation along the protein $\alpha$-helix.

The present study, while being focused on the effects of the symmetry parameter $\xi$ on the quantum dynamics of Davydov solitons, is limited in that it did not include a thermal bath. Consequently, the reported soliton lifetime of over 100~ps should not be directly extrapolated to physiological temperatures of $T=310$~K. Previous experimental studies have reported self-trapped amide~I states in peptide model crystals \cite{Edler2002} or protein (myoglobin) \cite{Xie2000} for about 15 ps. Computational studies using a quantum Monte Carlo method have found Davydov solitons to be unstable at temperatures above $T=7$~K \cite{Wang1989}, whereas studies using a thermally averaged Hamiltonian \cite{Cruzeiro1988} or a Langevin term obeying the fluctuation-dissipation theorem applied to the lattice \cite{Forner1997a,Forner1997b} have shown that for a certain range of parameters, Davydov solitons could persist at $T=310$~K for tens of picoseconds. In view of this controversy, F\"{o}rner \cite{Forner1997a} has pointed out that increasing the spring constant of the hydrogen bonds $w$ is one of several promising ways to provide thermal stability of Davydov solitons because the usually used value of $w=13$ N/m is measured in crystalline formamide \cite{Itoh1972,Scott1984,Scott1985}, where the formamide molecules can vibrate freely against each other in the potential due to the hydrogen bonds, while in an $\alpha$-helix on the contrary they are bound covalently in the backbone of the protein. Another mechanism for enhancement of thermal stability of Davydov solitons is to increase the number of amide~I quanta to at least two, which corresponds to the amount of energy released by a single ATP molecule \cite{Cruzeiro1994,Forner1997a}. Ab initio quantum chemical calculations of different parameters for the $\alpha$-helix and quantum-mechanical methods that introduce the effect of thermal agitation on amide I propagation, deserve further exploration.

\section*{Conflict of interest}

The authors declare that they have no conflict of interest.

\section*{Acknowledgment}

{We would like to thank the two anonymous reviewers for providing useful suggestions for improving the presentation of this work.}

\section{Supplementary videos}

\paragraph{Video~1}
Exciton dispersion at $\bar{\chi}=10$~pN for the case shown in Fig.~\ref{fig:3}a with symmetric Hamiltonian $\xi=1$
and an initial Gaussian step distribution $\sigma$ of amide~I energy over 3 peptide groups starting from the N-end
of an $\alpha$-helix spine composed of 40 peptide groups (extending along the $x$-axis) during a period
of 100 ps. Quantum probabilities $|a_n|^2$ are plotted in blue along the $z$-axis. Phonon lattice displacement differences $b_n-b_{n-1}$ (measured in picometers) are plotted in red along the $y$-axis.

\paragraph{Video~2}
Moving soliton at $\bar{\chi}=30$~pN for the case shown in Fig.~\ref{fig:3}b with symmetric Hamiltonian $\xi=1$
and an initial Gaussian step distribution $\sigma$ of amide~I energy over 3 peptide groups starting from the N-end
of an $\alpha$-helix spine composed of 40 peptide groups (extending along the $x$-axis) during a period
of 100 ps. Quantum probabilities $|a_n|^2$ are plotted in blue along the $z$-axis. Phonon lattice displacement differences $b_n-b_{n-1}$ (measured in picometers) are plotted in red along the $y$-axis. The soliton is formed by self-trapping of the amide~I energy by the induced lattice distortion.

\paragraph{Video~3}
Moving soliton at $\bar{\chi}=50$~pN for the case shown in Fig.~\ref{fig:3}c with symmetric Hamiltonian $\xi=1$
and an initial Gaussian step distribution $\sigma$ of amide~I energy over 3 peptide groups starting from the N-end
of an $\alpha$-helix spine composed of 40 peptide groups (extending along the $x$-axis) during a period
of 100 ps. Quantum probabilities $|a_n|^2$ are plotted in blue along the $z$-axis. Phonon lattice displacement differences $b_n-b_{n-1}$ (measured in picometers) are plotted in red along the $y$-axis. The soliton is formed by self-trapping of the amide~I energy by the induced lattice distortion.

\paragraph{Video~4}
Pinned soliton at $\bar{\chi}=70$~pN for the case shown in Fig.~\ref{fig:3}d with symmetric Hamiltonian $\xi=1$
and an initial Gaussian step distribution $\sigma$ of amide~I energy over 3 peptide groups starting from the N-end
of an $\alpha$-helix spine composed of 40 peptide groups (extending along the $x$-axis) during a period
of 100 ps. Quantum probabilities $|a_n|^2$ are plotted in blue along the $z$-axis. Phonon lattice displacement differences $b_n-b_{n-1}$ (measured in picometers) are plotted in red along the $y$-axis. The soliton is formed by self-trapping of the amide~I energy by the induced lattice distortion.

\appendix

\section{The generalized Ehrenfest theorem}
\label{appA}

{The generalized Ehrenfest theorem \cite{Lippmann1966} follows directly from the Schr\"{o}dinger equation and represents the underlying quantum dynamics.}
Suppose that we have an observable $\hat{A}(x,t)$ and a quantum system
in a state $|\Psi(x,t)\rangle$ evolving in time.
After differentiation, we have
\begin{equation}
\frac{d}{dt}\langle A(x,t)\rangle = \left(\frac{\partial}{\partial t}\langle\Psi|\right)\hat{A}(x,t)|\Psi\rangle
+\langle\Psi|\left(\frac{\partial}{\partial t}\hat{A}(x,t)\right)|\Psi\rangle
+\langle\Psi|\hat{A}(x,t)\left(\frac{\partial}{\partial t}|\Psi\rangle\right)\label{eq:dA-dt}
\end{equation}
The Schr\"{o}dinger equation and its complex conjugate transpose are
\begin{equation}
\frac{\partial}{\partial t}|\Psi\rangle = -\frac{\imath}{\hbar}\hat{H}|\Psi\rangle ;\quad\quad
\frac{\partial}{\partial t}\langle\Psi| = \frac{\imath}{\hbar}\langle\Psi|\hat{H}
\end{equation}
where we used the Hermiticity of the Hamiltonian
$\hat{H}=\hat{H}^{\dagger}$. Substituting these equations in \eqref{eq:dA-dt}
gives
\begin{equation}
\frac{d}{dt}\langle A(x,t)\rangle 
= \frac{1}{\imath\hbar}\langle\left[\hat{A}(x,t),\hat{H}\right]\rangle+\langle\frac{\partial}{\partial t}\hat{A}(x,t)\rangle\label{eq:gen-Ehrenfest}
\end{equation}
where $\left[\hat{A}(x,t),\hat{H}\right]=\hat{A}(x,t)\hat{H}-\hat{H}\hat{A}(x,t)$
is the commutator of $\hat{A}(x,t)$ with the Hamiltonian~$\hat{H}$.
For time independent operator $\hat{A}$, the expectation value of
its partial time derivative is zero, $\langle\frac{\partial}{\partial t}\hat{A}\rangle=0$,
hence the last term in Eq.~\ref{eq:gen-Ehrenfest} disappears.

\section{Deviation vector of the \texorpdfstring{$|D_{2}\rangle$}{D2} ansatz}
\label{appB}

The $D_2$ theorem by Brown \cite{Brown1988} states that Davydov's $|D_2(t)\rangle$ ansatz satisfies the Schr\"{o}dinger equation of the Fr\"{o}hlich Hamiltonian \cite{Frohlich1952,Frohlich1954} if and only if $\chi=0$. Thus, there will be some non-zero error when using $|D_{2}\rangle$ ansatz to study the quantum dynamics of Davydov solitons in proteins, however, Brown's theorem does not provide a measure of the deviation of the $|D_{2}\rangle$ ansatz from the true Schr\"{o}dinger solution. To quantify the deviation from the Schr\"{o}dinger equation,
we follow the method by Sun \emph{et al.} \cite{Sun2010} and calculate the amplitude of the deviation vector $|\delta(t)\rangle$,
which is defined as
\begin{equation}
\Delta(t)\equiv\sqrt{\langle\delta(t)|\delta(t)\rangle}
\end{equation}
where
\begin{equation}
|\delta(t)\rangle\equiv\imath\hbar\frac{\partial}{\partial t}|D_{2}(t)\rangle-\hat{H}|D_{2}(t)\rangle
\end{equation}
Because we have transformed away certain terms $\gamma(t)$ contributing a highly oscillatory pure phase to the dynamic Davydov equations,
the deviation vector of the $|D_{2}\rangle$ ansatz is reduced to
\begin{equation}
|\delta(t)\rangle=\sum_{n}\left[\imath\hbar\frac{da_{n}}{dt}+J\left(a_{n+1}+a_{n-1}\right)-\chi a_{n}\left(\hat{u}_{n+1}+(\xi-1)\hat{u}_{n}-\xi\hat{u}_{n-1}\right)\right]\hat{a}_{n}^{\dagger}|0_{\textrm{ex}}\rangle|b\rangle
\end{equation}
Taking the inner product further gives
\begin{eqnarray}
\langle\delta(t)|\delta(t)\rangle & = & \sum_{n}\Bigg[\hbar^{2}\left|\frac{da_{n}}{dt}\right|^{2}
+2\hbar J\,\textrm{Im}\left[\frac{da_{n}^{*}}{dt}\left(a_{n+1}+a_{n-1}\right)\right]
+J^{2}\left|a_{n+1}+a_{n-1}\right|^{2}\nonumber \\
& & +2 \left\{ \hbar\chi\textrm{Im}\left[\frac{da_{n}}{dt}a_{n}^{*}\right]
-\chi J\,\textrm{Re}\left[a_{n}^{*}\left(a_{n+1}+a_{n-1}\right)\right] \right\} 
\left(b_{n+1}+(\xi-1)b_{n}-\xi b_{n-1}\right) \nonumber \\
& & +\chi^{2}\left|a_{n}\right|^{2}\left(b_{n+1}+(\xi-1)b_{n}-\xi b_{n-1}\right)^{2}\Bigg]
\end{eqnarray}
In our simulations, the error for deviation from the Schr\"{o}dinger equation is found to be non-zero (Fig.~\ref{fig:9}),
but negligible in comparison with the soliton energy $E_{\textrm{sol}}=\langle D_{2}(t)|\hat{H}|D_{2}(t)\rangle$, which is
\begin{eqnarray}
E_{\textrm{sol}}	& = &	E_0 + W_0 -\sum_{n}\Bigg\{2J\left[\textrm{Re}(a_{n})\textrm{Re}(a_{n+1})+\textrm{Im}(a_{n})\textrm{Im}(a_{n+1})\right]
-\chi|a_{n}|^{2}\left(b_{n+1}+(\xi-1)b_{n}-\xi b_{n-1}\right)\nonumber\\
		&&\qquad\qquad\qquad\qquad-\frac{1}{2}M\left(\frac{db_{n}}{dt}\right)^{2} -\frac{1}{2}w\left(b_{n+1}-b_{n}\right)^{2}\Bigg\} \label{eq:solE}
\end{eqnarray}
where $W_0$ is the zero-point energy of the lattice vibrations. In the absence of a thermal bath, $E_{\textrm{sol}}$ corresponds to the total energy of the system and is a conserved quantity.
Because the contribution from the two constant energies $E_0 + W_0$ is two orders of magnitude larger than the remaining energy terms, in the main text we have reported the gauge transformed soliton energy $E = E_{\textrm{sol}} - E_0 - W_0$.
The error is negligible, $\Delta(t)<10^{-5} |E|$, for the whole dynamic range of $\bar{\chi}\leq70$~pN including the threshold for soliton pinning, which shows that the Davydov soliton is an excellent approximation to the full quantum dynamics in the absence of thermal agitation.
Thus, the $|D_{2}\rangle$ ansatz is dynamically equivalent to the mixed quantum-classical approach in which the lattice is considered classical \cite{Kenkre1994,Cruzeiro1997}.

\bibliography{references}

\end{document}